\documentclass[aps,prd,preprint,tightenlines,nofootinbib,showpacs]{revtex4}
\usepackage{epsfig,afterpage}
\usepackage{colordvi}

\newcommand{\etal}{{\it et al.}}
\newcommand{\gev}{{\ifmmode{{{\text{GeV}}}}\else{{\mbox{GeV}}}}}

\begin{document}


\preprint{SU-HEP-010}
\vspace*{1cm}
\title{\boldmath S-waves and the Measurement of CP Violating Phases in $B_s$ Decays}

\author{S.~Stone}
\author{L.~Zhang}
\affiliation{Physics Department, Syracuse University, Syracuse, New York 13244}

\date{December 18, 2008}

\begin{abstract}
Heavy, as yet undiscovered particles, can affect measurements of CP violation in the $B$ system.
Measuring CP violation in the $B_s$ system provides an excellent place to observe such effects since Standard Model sources are predicted to produce very small effects. The angle
$-2\beta_s$, the ``phase of $B_s-\overline{B}_s$ mixing," thought to be best measured in $B_s\to J/\psi\phi$ decays is of order -0.04, while the CP violating asymmetry in $B_s\to\phi\phi$ is predicted to be zero, due to the cancellation of the mixing phase with the decay phase. Recent measurements of $\beta_s$ in $J/\psi\phi$, while not definitive, are much larger than the Standard Model predictions. Measurements in the $B^0$ and $D_s^+$ systems of analogous modes point toward a 5-10\% contamination of S-wave $K^+K^-$ under the $\phi$ peak. This S-wave was not taken into account in these recent analyses. Furthermore this S-wave can also materialize as a $f_0(980)$ meson that decays to $\pi^+\pi^-$, making the final state $J/\psi f_0$ useful for measuring $\beta_s$ with the added advantage of not requiring an angular analysis. Rate estimates, while not precise, predict four to five times fewer such events than those in the $J/\psi\phi$ mode. The error on $\beta_s$, however, may be similar. We also remark on S-wave problems with the $B_s\to\phi\phi$ mode, and possible systematic checks using $B_s\to \phi f_0$.
\end{abstract}

\maketitle \tighten


Measurements of Charge-Parity (CP) violation in the $B$ meson system are sensitive to the presence of heavy, as yet undiscovered, particles. While CP violation has been studied extensively in the $B^0$ system \cite{Fleischer}, where it is a large effect, very little information exists for the $B_s$ system. The CP violating angle $-2\beta_s$, the so called ``phase of $B_s-\overline{B}_s$ mixing" is a particularly important place to look for physics beyond the Standard Model, since the expected asymmetry is very small,  $\sin(2\beta_s)=0.037\pm 0.002$ \cite{Charles}, thus allowing the effects of any new physics to be more easily observed.
Both CDF \cite{CDF-chi} and D0 \cite{D0-chi} have investigated $-2\beta_s$ using $B_s\to J/\psi\phi$ decays. Central values have been found far from the expected Standard Model values, but the errors are large and the significance is in the 2-3$\sigma$ range.

Since the final state consists of two spin-1 particles, the final state is not a CP-eigenstate, yet it is well known that CP violation can be measured using angular analyses \cite{transversity}. The analysis that have been heretofore carried out have ignored the possibility of an S-wave $K^+K^-$ system in the region of the $\phi$. This S-wave can bias the result and not accounting for it certainly makes the error smaller. In fact, the analogous channel in $B_d$ decay $J/\psi K^{*0}$ is well known to have an S-wave $K\pi$ component in the $K^*$ mass region. This interference, in fact, has been used by BaBar to measure $\cos(2\beta)$ and thus remove an ambiguity in the value of $\beta$ from the $\sin(2\beta)$ measurement. The S-wave component in the region of the $K^*$ is measured as $\approx8$\% \cite{Babar-psiKstar}.

\begin{figure}[htb]
  \begin{center}
    \leavevmode
     \epsfxsize=.65 \textwidth
     \hskip 0in \epsfbox{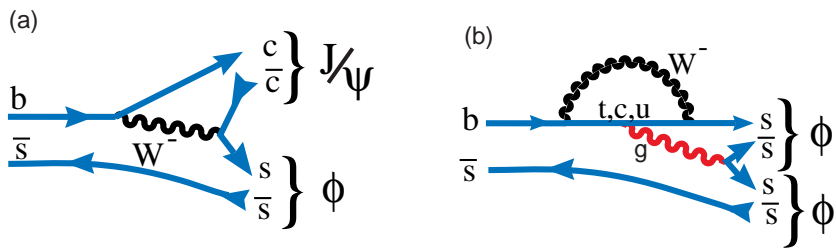}
  \end{center}
\vskip -0.15in
\caption{Decay diagrams (a) $B_s\to J/\psi \phi$, and (b) $B_s\to \phi\phi$.}
  \label{psi_ss}
\end{figure}

Perhaps it may be hoped that the S-wave $K^+K^-$ under the $\phi$ in $J/\psi\phi$ is smaller due to the relatively narrow width  ($\Gamma$) of the $\phi$ (4.3 MeV) compared to the $K^*$ (51 MeV), but we will show here that this likely is not the case. (The S-wave problem was brought up at the recent CKM workshop in the discussion of the CDF and D0 results \cite{Yuehong}.) This adds another amplitude and phase to the number of parameters that need to be determined in the full transversity analysis and increases the error over previous expectations \cite{Gaia}.
Similar considerations apply to the measurement of CP violation in the process $B_s\to \phi\phi$. Here the problem is exacerbated by the presence of two $\phi$'s in the final state. The decay diagrams for both of these processes are shown in Fig.~\ref{psi_ss}. In both cases the $s\overline{s}$ forms a $\phi$. Other manifestations of $s\overline{s}$ quarks are the $\eta$, $\eta'$ and $f_0$(980) mesons. The first two are pseudoscalars, while the last is a scalar.

Final states with $J/\psi$ plus a spin-0 object have the $J/\psi$ fully polarized in the ($J,~J_z$) (1,0) state, and thus an angular analysis is unnecessary. Unfortunately, the measurement of $\beta_s$ using the pseudoscalars has less sensitivity than using the $\phi$
at hadron colliders, because the large decay modes of the $\eta$ contain at least two photons and the $\eta'$ at least one.  Due to the relatively poor photon detection efficiency,  larger backgrounds, and poor mass resolutions using photons at current experiments, the accuracy on $\beta_s$ is much poorer using the $\eta$ or $\eta'$ modes than in the $\phi$ mode, even though the angular analysis can be avoided. The scalar $f_0$ state, however, hasn't been previously considered.

\begin{figure}[htb]
  \begin{center}
    \leavevmode
     \epsfxsize=.65 \textwidth
     \hskip 0in \epsfbox{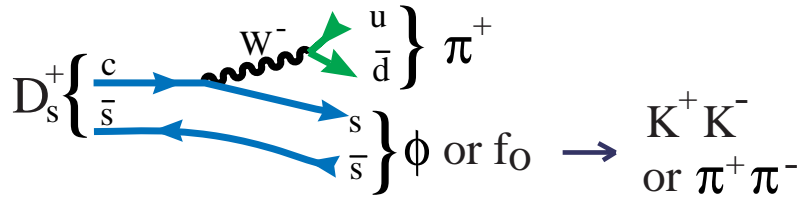}
  \end{center}
\caption{Decay diagram for a $D_s^+$ decay into a $\pi^+$ plus an $s\overline{s}$ system.}
  \label{Dsdecay}
\end{figure}

We note that the $f_0(980)$ is an elusive object that decays largely into $\pi^+\pi^-$ but can also decay into $K^+K^-$. Some information about this particle can be gleaned from analysis of exclusive $D_s^+$ final states. The $\pi^+\pi^+\pi^-$ and $K^+K^-\pi^+$ are of prime interest.
The simple spectator decay diagram that results in the $\phi\pi^+$ or the $f_0\pi^+$ final states is shown in Fig.~\ref{Dsdecay}. We note that the $s\overline{s}$ system can form only isoscalar final states.

\begin{figure}
\centerline{\epsfig{figure= 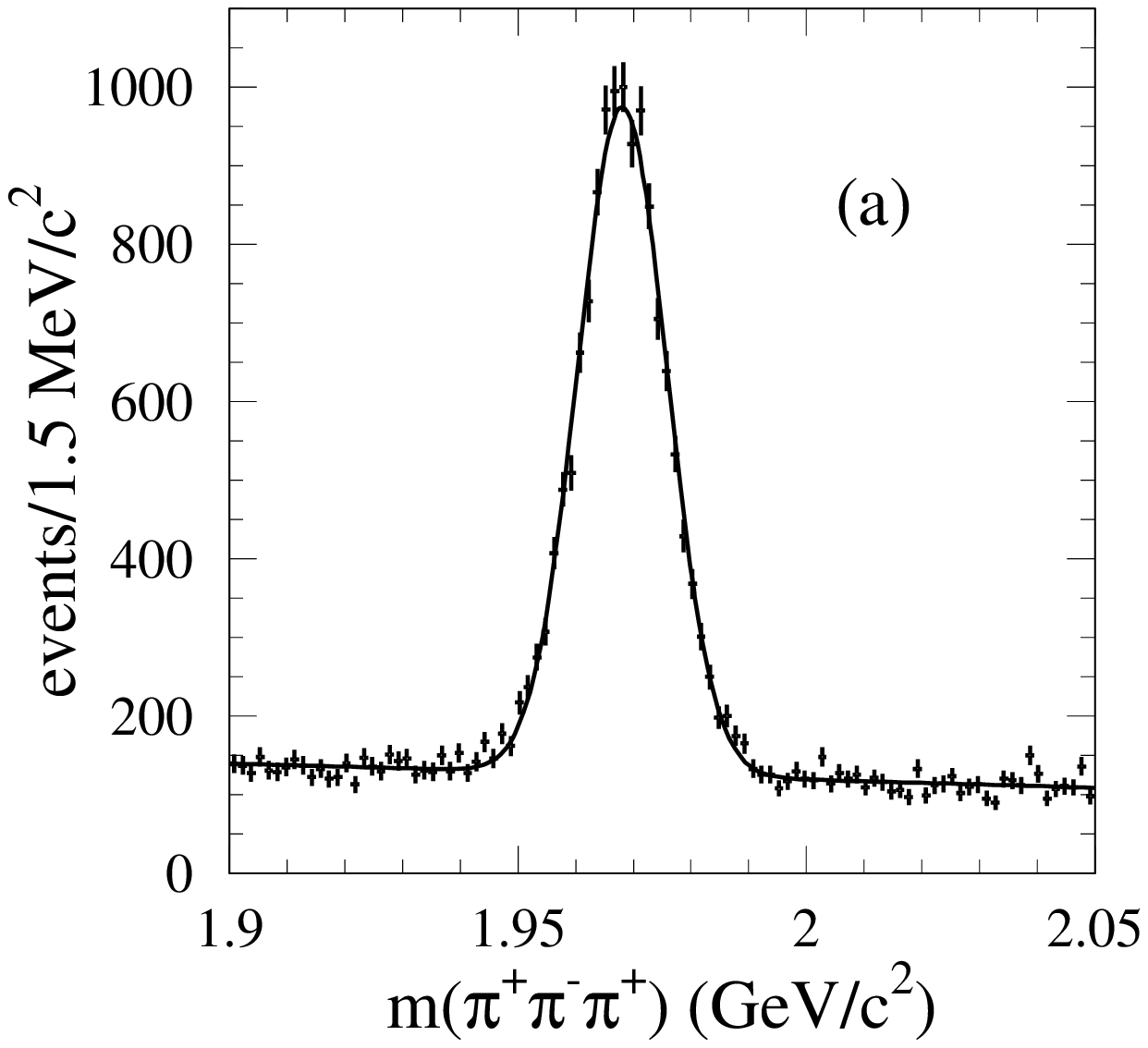,height=2.1in}\hspace{1cm}\epsfig{figure= 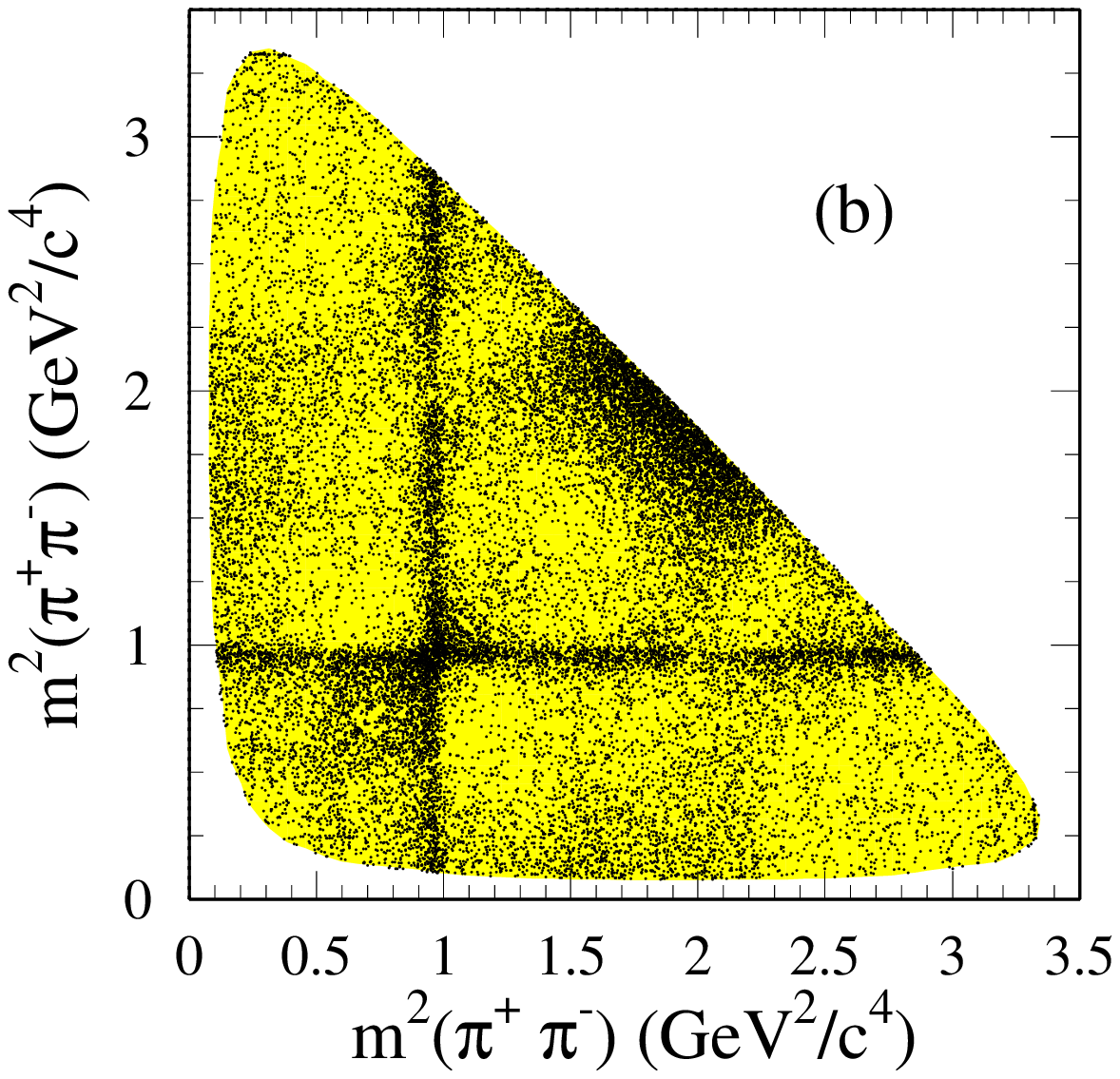,height=2.1in}}
\caption{(a) The $\pi^+\pi^+\pi^-$ invariant mass distribution.
(b) The symmetrized $D^+_s \to \pi^+\pi^+\pi^-$ Dalitz plot. From the
BaBar collaboration \cite{Babar-Dalitz}.}
\label{3piDalitz}       
\end{figure}

Of course, not only these final states are produced. A full Dalitz plot analysis, however, reveals the structure. The most recent and highest statistics study
of the $\pi^+\pi^+\pi^-$ final state has be done by BaBar \cite{Babar-Dalitz}. The invariant mass spectrum and Dalitz plot are shown in Fig.~\ref{3piDalitz}.
There best fit results showing fractions and phases
are summarized in Table~\ref{tab:fit_pipipi}.
\begin{table}[!htb]
\centering
\begin{tabular}{cccc}
\hline
Decay Mode\rule{0pt}{9pt}  & Decay fraction(\%)          & Amplitude                       & Phase(radians)   \\
\hline
\hline
$f_2(1270)\pi^+$\rule{0pt}{9pt}      & $10.1 \pm 1.5\pm 1.0 $             & 1.(fixed)                       & 0.(fixed) \\
$\rho(770)\pi^+$      & $1.8  \pm 0.5\pm1.0 $             & $0.19 \pm 0.02\pm0.12$                 & $1.1 \pm 0.1\pm0.2$\\
$\rho(1450)\pi^+$     & $2.3  \pm 0.8\pm1.7 $             & $1.2 \pm 0.3\pm1.0$                   & $4.1 \pm 0.2\pm 0.5$\\
{S}-wave    & $83.0 \pm 0.9\pm 1.9 $             &  & \\
\hline
TOT.                  & $97.2 \pm 3.7 \pm3.8$             &                                 & \\
$\chi^2/NDF$          &  $\frac{437}{422-64}$ = 1.2 &                                 &\\
\hline
\end{tabular}
\caption{Results from the BaBar $D_s^+ \to \pi^+ \pi^- \pi^+$ Dalitz plot analysis. The table reports the fit fractions, amplitudes and
phases. Errors are statistical and systematic, respectively.}
\label{tab:fit_pipipi}
\end{table}

\begin{figure}
\centerline{\epsfig{figure=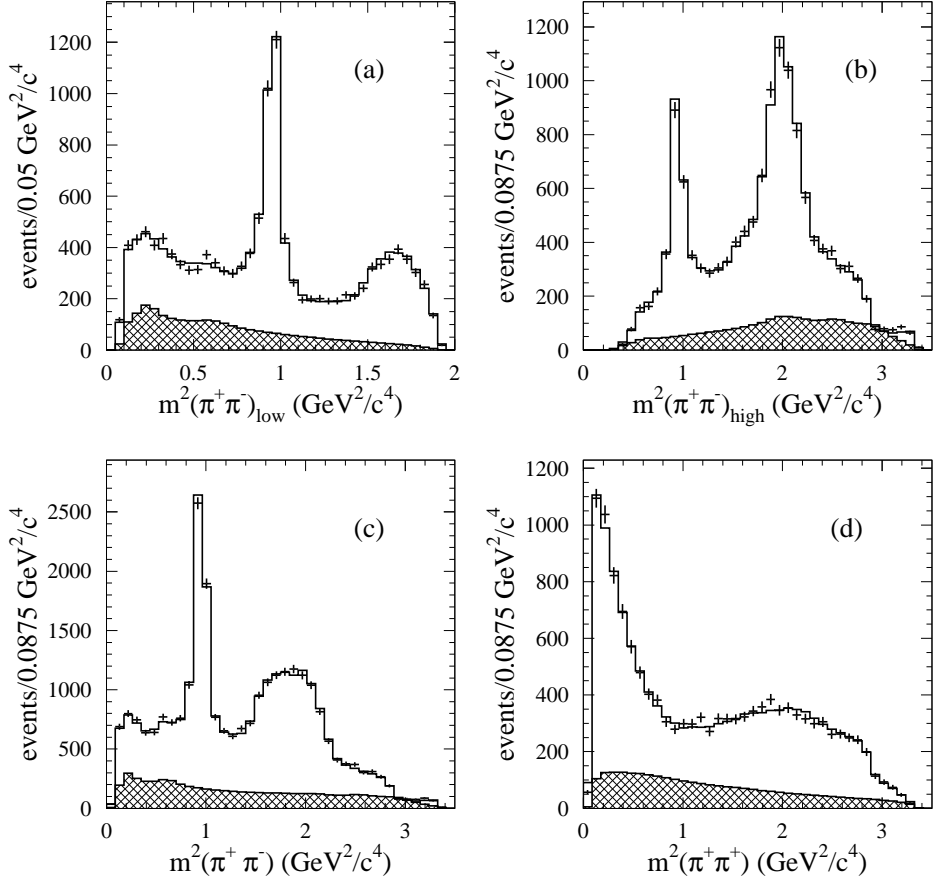,height=4.9in}}
\caption{Dalitz plot projections (dots with error bars) and fit results (solid histogram).
(a) $m^2(\pi^+ \pi^-)_{Low}$, (b) $m^2(\pi^+ \pi^-)_{High}$, (c) total $m^2(\pi^+ \pi^-)$, (d) $m^2(\pi^+ \pi^+)$.
The hatched histograms show the background distribution. From the BaBar Collaboration \cite{Babar-Dalitz}.}
\label{pipimass}       
\end{figure}

Thus, most of this final state is S-wave (83\%). The mass projection of the Dalitz plot shown in Fig.~\ref{pipimass} shows a relatively narrow peak in $\pi^+\pi^-$ mass within $\sim\pm$90 MeV of the $f_0$ mass. This is in agreement with a previous analysis by FOCUS \cite{Focus}\cite{E791}.

A Dalitz plot analyses of the $K^+K^-\pi^+$ final state have been performed by E687 \cite{E687}. They find the results shown in Table~\ref{tab:fit_KKpi}.
\begin{table}[!htb]
\centering
\begin{tabular}{cccc}
\hline
Decay Mode\rule{0pt}{9pt}  & Decay fraction(\%)                          & Phase(degrees)   \\
\hline
\hline
$\overline{K}^*(892)^0K^+$ & $0.478\pm 0.046\pm 0.040$ & 0 (fixed)  \\
$\phi\pi^+$ & $0.396\pm0.033\pm0.047$ & $178\pm20\pm 24$ \\
$f_0(980)\pi^+$ & $0.110\pm0.035\pm0.026$ & $159\pm22\pm 16$ \\
$f_J(1710)\pi^+$ & $0.034\pm0.023\pm0.035$ & $110\pm20\pm 17$ \\
$\overline{K}^*(1430)^0K^+$ & $0.093\pm 0.032\pm 0.032$ & 0 $152\pm 40\pm 39$  \\
\hline
Goodness of Fit          &  -2ln${\cal{L}}$=-1075 &   conf. level=80.2\%                      \\
\hline
\end{tabular}
\caption{Results from the E687 $D_s^+ \to K^+K^- \pi^+$ Dalitz plot analysis. The table reports the fit fractions, and
phases.}
\label{tab:fit_KKpi}
\end{table}
While extracting a precise ratio of decay rates is difficult, because the phases of the $f_0\pi^+$ and $\phi\pi^+$ amplitudes are almost equal, within error, we can infer that the ratio is
\begin{equation}
\frac{\Gamma(D_s^+\to f_0(980)\pi^+\to K^+ K^-\pi^+)}{\Gamma(D_s^+\to\phi\pi^+\to K^+ K^-\pi^+)}
=0.28\pm 0.12~~.
\end{equation}
Clearly one cannot ignore the S-wave contribution here. However, this analysis is done over all of phase space and we concerned only with the low mass region.

\begin{figure}
\centerline{\epsfig{figure=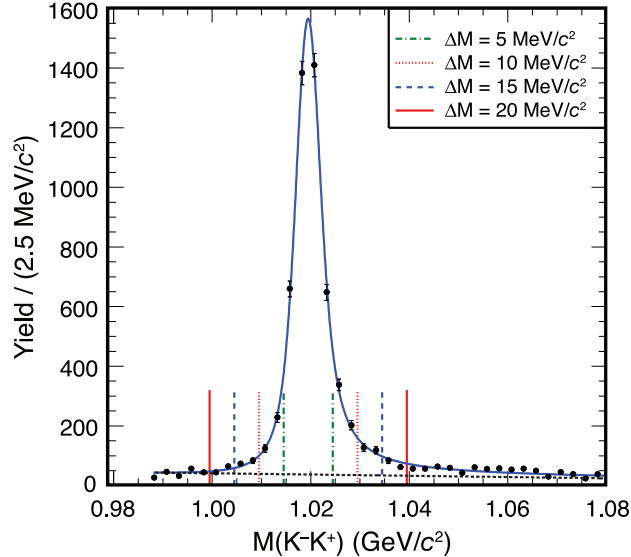,height=2.9in}}
\caption{Dalitz plot projections for $K^+K^-$ invariant mass in $D_s^+\to K^+ K^-\pi^+$ from the CLEO Collaboration \cite{CLEOcabs}. The signal is extracted individually in each mass bin, thus there is no background. The data are fit with a Breit-Wigner signal function for the $\phi$ convoluted with a Gaussian for detector resolution and linear representation of an S-wave component (dashed line). The solid curve shows the sum. (Only the data is ascribed to CLEO, the fits have been added.)}
\label{CLEOmKK}       
\end{figure}

CLEO has looked explicitly at the low mass $K^+K^-$ region in the $K^+ K^-\pi^+$ Dalitz plot \cite{CLEOcabs}. Fig.~\ref{CLEOmKK} shows the data in the region near 1 GeV. The signal is extracted by fitting the $D_s$ yield in each bin of $K^+K^-$, so no background remains in the plot. There is clearly an extra rather flat component of signal beneath the $\phi$. To estimate the size of this component, we have fit the CLEO data to a Breit-Wigner to describe the $\phi$, convoluted with a Gaussian for detector resolution, and in addition added a linear S-wave component. The fraction of S-wave depends on the mass interval considered. For $\pm$10 MeV around the $\phi$ mass we have 6.3\% S-wave contribution, which rises to 8.9\% for a $\pm$15 MeV interval. (Note that these fractions depend on the experimental resolution.)

 We now try to compare the S-wave in $D_s$ decays with the one in $B_s\to J/\psi K^+K^-$. The near equality of the $f_0$ and $\phi$ masses removes this from consideration. Let us consider the energy. In the case of $D_s\to K^+K^- \pi^+$ we have 1.97 GeV minus the pion mass, or 1.83 GeV available. In the case of $B_s$ we have 5.37 GeV minus 3.10 GeV or 2.27 GeV. A bit more energy but not that different. What does differ is that we have a $0^-$ pion in one case and a $1^-$ $J/\psi$ in the other case, and the spin may matter.

Now we wish to predict how many events we may get should we look for $B_s\to J/\psi f_0$ with $f_0\to\pi^+\pi^-$.
CLEO-c measures ${\cal{B}}(D_s^+\to K^+K^-\pi^+)=(5.50\pm0.23\pm0.16)$\% \cite{CLEOcabs}. First we need to derive a $\phi\pi^+$ branching ratio (with $\phi\to K^+K^-$).
CLEO-c gives a range of branching ratios computed for varying cuts around the $\phi$ mass. Extrapolating their data to a zero width interval allows us to remove the S-wave component under the $\phi$ (see Fig.~\ref{CLEOmKK}). This gives a value of ${\cal{B}}(D_s^+\to \phi\pi^+,$ $\phi\to K^+K^-)$=(1.6$\pm$0.1)\%.
 We also use the CLEO absolute branching fraction measurement ${\cal{B}}(D_s^+\to \pi^+\pi^-\pi^+)=(1.11\pm 0.07\pm 0.04)$\%, and
 the BaBar Daltiz analysis of the 3$\pi$ mode to derive a branching fraction for ${\cal{B}}(D_s^+\to f_0\pi^+\to \pi^+\pi^+\pi^-)=0.30$\%, where we found by integrating that (27$\pm$2)\% of the 3$\pi$ mode is due to a narrow $f_0$ within $\pm 0.13$ GeV$^2$ of the $f_0$ mass-squared ($\approx\pm$90 MeV of the $f_0$ mass of 980 MeV).  This comes from an examination of the BaBar data shown in Fig.~\ref{pipimass}(c). Thus, we expect a sample of events $B_s\to J/\psi f_0,~f_0\to\pi^+\pi^-$ that is (19$\pm$2)\% of the $J/\psi \phi$ sample.
 If we include the contribution of all the $\pi^+\pi^-$ S-wave combinations above
background in the three peak bins, our estimate would increase to (27$\pm$3)\%, as there is significant $\pi^+\pi^-$ S-wave
outside of the $f_0$ mass region.

We now make another estimate of the relative $f_0\to\pi^+\pi^-/\phi\to K^+K^-$ rate. Semileptonic $D_s$ decays provide another basis of estimate. Here the $s\overline{s}$ pair is produced opposite a virtual $W^-$, and thus is produced opposite an object with the same spin as the $J/\psi$. Measuring the ratio of semileptonic decay rates where the final state meson has the maximum momentum (at $q^2=0$, where $q^2$ is the 4-momentum transfer) would approximate the available energy quite well (see above).  CLEO has recently measured the branching ratios integrated overall $q^2$ \cite{CLEO-semi}
\begin{eqnarray}
{\cal{B}}(D_s^+\to \phi e^+\nu),~\phi\to K^+ K^- &=& (1.1\pm 0.2)\% \\\nonumber
{\cal{B}}(D_s^+\to f_0 e^+\nu),~f_0\to\pi^+\pi^- &=& (0.13\pm 0.04)\%~. \\
\end{eqnarray}
Unfortunately, the $q^2$ distributions are not yet available so we will use the integrated rates; we know the maximum rate is at $q^2$ of zero. The ratio of observable $f_0/\phi$ here is (13$\pm$4)\%, where the error is only from the CLEO data and not the model approximations. This estimate is very qualitative in that we do not have the $q^2$ equals zero data, but it is consistent with our previous estimate, given the limitations in the data.

Thus, using our value from hadronic $D_s$ decays, we expect that
\begin{equation}
\frac{\Gamma(B_s^0\to J/\psi f_0,~f_0\to \pi^+\pi^-)}{\Gamma(B_s^0\to J/\psi \phi,~\phi\to K^+K^-)}\approx 20\%~.
\end{equation}

This rate may increase to 30\% if there is additional S-wave under the $f_0$. We note that any $\pi^+\pi^-$ S-wave, or even D-wave, does not constitute a background for the CP violation measurement as it has the same CP as the $f_0$; also, since the $\rho^0$ doesn't have any $s\bar{s}$ in its wave-function, it is unlikely that the $s\bar{s}$ system forms a $1^-$ state \cite{rho}. The purity of the sample can be checked by examining the polarization of the leptons from the $J/\psi\to \ell^+\ell^-$ decay; this serves as an experimental check on the amount of opposite CP P-wave. For pure $f_0$, in the limit of massless leptons the angle $\theta$ of the $\ell^+$ in the $J/\psi$ rest frame with respect to the $f_0$ direction in this frame, must be distributed as $\sin^2\theta$.

 In the decay $B_s\to \phi\phi$, the decay phase cancels the mixing phase if only Standard Model particles are involved, so this reaction provides an interesting place to look for manifestations of new physics \cite{Mohanty}.
 The presence of the S-wave in the $K^+K^-$ distributions at low mass also applies to the $\phi\phi$ final state. Here, however, there are two possible S-wave contributions that must be taken into account since there are two $\phi$'s.  It may be that using the $\phi f_0$ final state can help in understanding this effect. Here an angular analysis is only necessary to disentangle the amount of S-wave under the $\phi$ peak. We note however, that events with two charged kaons and two charged pions can also result from the $\overline{K}^{*0}K^{*0}$ final state, so a great deal of care must be exercised. (These final states seem to be well separated by the large phase space available, at least at first glance.)

 In conclusion, we predict on the order of a 10\% $K^+K^-$ S-wave contribution that contaminates
 $B_s^0\to J/\psi \phi,~\phi\to K^+K^-$ and must be taken into account by adding an additional S-wave amplitude and phase to the parameterization of the decay width. Inclusion of this S-wave will change the central values and increase the errors in current analyses of $\beta_s$. We suggest a new decay
 $B_s\to J/\psi f_0,~f_0\to\pi^+\pi^-$. Rough estimates indicate that we can expect at least $\sim$1/5 as many events as in $J/\psi\phi$.  An angular analysis is not necessary to find $\beta_s$ using these events, and thus the statistical error on $\beta_s$ could be comparable to the and the systematic error smaller than those for the $J/\psi\phi$ mode.  A critical look needs to be given towards the angular analysis of the $B_s\to \phi\phi$ mode since the S-wave can enter twice. It may be possible to access the same physics using the $B_s\to \phi f_0$, where an angular analysis may only be necessary to ascertain the fraction of S-wave under the $\phi$ peak. Measurement of CP violation in the $B_s$ system continues to be of utmost importance.

 We would like to thank M. Artuso, J. Rosner and several LHCb colleagues for useful discussions. We thanks the U. S. National Science Foundation for support.



\newpage


\begin{thebibliography}{99}
\bibitem{Fleischer}
R. Fleischer, ``CP Violation and B Physics at the LHC," ECONFC0610161, 020 (2006)
[arXiv:hep-ph/0703112v2]; M. Artuso, E. Barberio and S. Stone, PMC Physics A, {\bf 3}:3 (2009), arXiv:0902.3743 [hep-ph].


\bibitem{Charles}
J. Charles \etal, Eur. Phys. J. C{\bf 41}, 1 (2005),
hep-ph/0406184.


\bibitem{CDF-chi}
T. Aaltonen \etal~(CDF), Phys. Rev. Lett. {\bf 100}, 161802 (2008),
arXiv:0712.2397v1 [hep-ex].

\bibitem{D0-chi}
V. M. Abazov \etal~(D0), Phys. Rev. Lett. {\bf 101}, 241801 (2008), arXiv:0802.2255v1 [hep-ex].

\bibitem{transversity}
I. Dunietz, H. R. Quinn, A. Snyder, W. Toki and H. J. Lipkin,
    Phys. Rev. D {\bf 43}, 2193 (1991); J. L. Rosner, Phys. Rev. D {\bf 42}, 3732 (1990);
 A. S. Dighe, I. Dunietz, H. J. Lipkin and J. L. Rosner, Phys. Lett.
    B {\bf 369}, 144 (1996), hep-ph/9511363.

\bibitem{Babar-psiKstar}
B. Aubert \etal~(BaBar), Phys. Rev. D{\bf 76}, 031102 (2007),
arXiv:0704.0522v2 [hep-ex].

\bibitem{Yuehong}
Private communication from Yuehong Xie. See http://ckm2008.roma1.infn.it/ for presentations
at the workshop.

\bibitem{Gaia}
See for example
G. Lanfranchi, ``Search for New Physics in $B_s\to J/\psi\phi$ decay @ LHC," presented at
CKM Workshop, Rome, September, 2008,
http://lhcb-doc.web.cern.ch/lhcb-doc/presentations/conferencetalks/postscript/2008presentations/ckm2008\_gaia-v1.pdf~.




\bibitem{Babar-Dalitz}
B. Aubert \etal~(BaBar), Phys. Rev. D {\bf 79}, 032003 (2009), arXiv:0808.0971 [hep-ex].

\bibitem{Focus}
J. M. Link \etal~(FOCUS), Phys. Lett. B {\bf 585}, 200 (2004).

\bibitem{E791}
A smaller statistic sample was also analyzed by E791, see
E. M. Aitala \etal~(E791), Phys. Rev. Lett. {\bf 86}, 765 (2001).

\bibitem{E687}
P.L. Frabetti \etal~(E687) Phys. Lett. B {\bf 351}, 591 (1995).

\bibitem{CLEOcabs}
J. P. Alexander \etal~(CLEO), Phys. Rev.Lett. {\bf 100}, 161804 (2008), arXiv:0801.0680v2 [hep-ex]. (We are authors of this paper.)



\bibitem{CLEO-semi}
J. Yelton \etal~(CLEO), arXiv:0903.0601 [hep-ex].

\bibitem{rho}
The decay fraction in Table~\ref{tab:fit_pipipi} shows a small $\rho^0$ component in the vicinity of the $f_0$.



\bibitem{Mohanty}
R. Mohanta and A.K. Giri, Phys. Rev. D {\bf 76}, 075015 (2007) arXiv:0707.1234 [hep-ph];
M. Raidal, Phys. Rev. Lett. {\bf 89}, 231803 (2002);
Y.~H. Chen \etal, Phys. Rev. D {\bf 59}, 074003 (1999).
\end{thebibliography}
\end{document}